\documentstyle[prl,aps,epsf,epsfig,array]{revtex}

\newcommand{\bi}{\begin{itemize}}
\newcommand{\ei}{\end{itemize}}
\newcommand{\bc}{\begin{center}}
\newcommand{\ec}{\end{center}}

\newcommand{\be}{\begin{equation}}
\newcommand{\ee}{\end{equation}}

\newcommand{\bqn}{\begin{eqnarray}}
\newcommand{\eqn}{\end{eqnarray}}

\begin{document}

\title{One loop renormalization of soliton quantum mass corrections in 
(1+1)-dimensional scalar field theory models}

\author{G. Flores-Hidalgo \thanks{E-mail:gflores@cbpf.br}}

\address{{\it  Centro Brasileiro de Pesquisas Fisicas-CBPF,
Rua Dr. Xavier Sigaud 150, 22290-180 Rio de Janeiro, RJ, Brazil}}

\maketitle

\begin{abstract}
The bare one loop soliton quantum mass corrections can be expressed 
in two ways: as a sum over the zero-point energies of small oscillations 
around the classical configuration, or equivalently as the (Euclidean) 
effective action per unit time. In order to regularize the bare one loop 
quantum corrections (expressed as the sum over the zero-point energies) 
we subtract and add from it the tadpole graph that appear in the 
expansion of the effective action per unit time. The subtraction 
renders the one loop quantum corrections finite. Next, we use the 
renormalization prescription that the added tadpole graph cancels with
adequate counterterms, obtaining in this way a finite unambiguous 
expression for the one loop soliton quantum mass corrections. When we apply 
the method to the solitons of the sine-Gordon and $\phi^4$ kink models we 
obtain results that agree with known results. Finally we apply the method 
to compute the soliton quantum mass corrections in the recently introduced 
$\phi^2\cos^2\ln(\phi^2)$ model.

\vspace{0.34cm}
\noindent
PACS number(s): 11.10.Gh, 11.15.Kc, 11.27.+d

\end{abstract}
\vspace{1cm}

A renewed interest in the computation of quantum energies around classical 
configurations has recently arose. See for example 
\cite{chan,graham1,dunne,goldhaber,bordag,aragao,wimmer} and references therein. The methods used to approach the problem include the derivative expanssion method 
\cite{chan}, the scattering phase shift technique \cite{graham1}, the mode
regularization method \cite{goldhaber}, the zeta-function regularization
technique \cite{bordag} and also the dimensional regularization method 
\cite{wimmer}. In this letter I will give a 
very simple derivation of the one loop renormalized soliton quantum mass 
correction in 1+1 dimensional scalar field theory models, using the scattering 
phase shift technique. The approach used here differ from those given in Ref. 
\cite{graham1}. Consequently, it is obtained a formula, Eq. (\ref{17}), 
that at first sight is different from the one obtained in Ref. \cite{graham1}.

Let us start with a Lagrangean density
\begin{equation}
{\cal L}=\frac{1}{2}\partial_\mu\phi\partial^\mu\phi-U(\phi)\;,
\label{1}
\end{equation}
where $\mu=0,1$. When $U(\phi)$ has at least two degenerate trivial vacua
the classical equation of motion admits static finite energy
solutions $\phi_c$. These solutions are called solitons \cite{rajaraman}. 
After quantizing around one of these static solutions, it is showed that 
there is a quantum state corresponding to this static solution. This
state is called the soliton quantum state \cite{jackiw1}. The soliton
quantum state behaves as a particle and in particular, the first contribution
different from zero to their mass is given by the (classical) energy of the
static solution. The one loop (bare) soliton quantum mass correction is given 
by
\begin{equation}
\Delta M_{bare}=\frac{1}{2}\sum_n\omega_n-\frac{1}{2}\sum_k\omega_k^0\;,
\label{2}
\end{equation}
where $\omega_n$ and $\omega_k^0$ are given respectively from the eigenvalue 
equations,
\begin{equation}
\left[-\frac{d^2}{dx^2}+U''[\phi_c(x)]\right]\psi_n(x)=\omega_n^2\psi_n(x)
\label{3}
\end{equation}
and
\begin{equation}
\left[-\frac{d^2}{dx^2}+m^2\right]\psi_k(x)=(\omega_k^0)^2\psi_k(x)\;,
\label{4}
\end{equation}
where $m^2=U''[\phi_c(\pm\infty)]$ is the mass of the quantum fluctuations
around the trivial vacua (we restrict ourselves to the case in which 
$U''[\phi_c(\infty)]=U''[\phi_c(-\infty)]$). From Eq. (\ref{4}) it 
is easy to see that the soliton-free eigenfunctions are given by 
$\psi_k\propto {\rm e}^{ikx}$ with $\omega_k^0=\sqrt{k^2+m^2}$. In general 
the eigenvalues of Eq. (\ref{3}) fall in two sets, a finite discrete set, that
we denote by $\omega_i^2$ and a continuum set, that we denote by $\omega^2(q)$ 
ranging from $m^2$ to $\infty$. The continuous eigenfunctions behave 
asymptotically as $\propto {\rm e}^{iqx}$ with modes 
$\omega(q)=\sqrt{q^2+m^2}$, {\it i.e}, equal to the soliton-free modes
$\omega_k^0$. Naively we can expect that the sum in Eq. (\ref{2}) will be 
reduced to a sum over the discrete modes $\omega_i$ only. But this is not the 
case, since there is a difference between the distributions of $\omega(q)$
and $\omega_k^0$ over $q$ and $k$ respectively, {\it i.e}, they have different 
densities of states. We can divide the sum in Eq. (\ref{2}) into a sum over the 
discrete modes and a sum over the continuum modes,
\begin{equation}
\Delta M_{bare}=\frac{1}{2}\sum_i\omega_i+\frac{1}{2}\sum_q\omega(q)-
\frac{1}{2}\sum_k 
\omega_k^0\;.
\label{5}
\end{equation}
In order to write the continuous sum in the above equation in an integral 
form, we first enclose the system in a box of size $L$, apply periodic boundary
conditions and finally take the limit $L\to\infty$. For simplicity we start
with the simplest case in which $U''[\phi_c(x)]$ is a reflectionless 
potential (this is the case for example, in the sine-Gordon and $\phi^4$ 
kink models), then we generalize the result for other potentials.
For reflectionless potentials the asymptotic behavior of $\psi_q(x)$ is 
given by,
\begin{equation}
\psi_q(x)=\left\{\begin{array}{ll}
{\rm e}^{iqx} &x\to-\infty\\
{\rm e}^{iqx+i\delta(q)} &x\to\infty
\end{array}\right.\;,
\label{6}
\end{equation}
where $\delta(q)$ is the so called phase shift. Placing the system into a box
of size $L$ and imposing periodic boundary conditions for the free 
eigenfunctions $\psi_k\propto {\rm e}^{ikx}$, we have ${\rm e}^{-ikL/2}={\rm e}
^{ikL/2}$, from which we obtain $k_n=\frac{2\pi n}{L}$, $n=\pm 1,\pm 2$. In 
this case the free density of states is $1/(k_{n+1}-k_n)=L/(2\pi)$. On the 
other hand from Eq. (\ref{6}), imposing periodic boundary conditions we get,
\begin{eqnarray}
q_n&=&\frac{2\pi n}{L}-\frac{\delta(q_n)}{L}\nonumber\\
&=&k_n-\frac{\delta(k_n)}{L}+{\cal O}(L^{-2})\;,
\label{7}
\end{eqnarray}
where when passing to the second line we have solved iteratively in terms of 
$k_n$. In Eq. (\ref{7}) we disregarded terms of order higher than 
${\cal O}(L^{-1})$, since in the limit $L\to\infty$ their contributions to 
$\Delta M_{bare}$ vanish. Replacing Eq. (\ref{7}) in $\sqrt{q_n^2+m^2}$ and 
expanding up to
${\cal O}(L^{-2})$, we obtain
\begin{equation}
\sqrt{q_n^2+m^2}=\sqrt{k_n^2+m^2}-\frac{k_n\delta(k_n)}{L\sqrt{k_n^2+m^2}}
+{\cal O}(L^{-2})\;.
\label{8}
\end{equation}
Replacing Eq. (\ref{8}) in Eq. (\ref{5}), using the free density of states 
$\frac{L}{2\pi}$, going to the continuum limit $L\to\infty$ and integrating
by parts we obtain,
\begin{equation}
\Delta M_{bare}=\frac{1}{2}\sum_i\omega_i-\left.\frac{\omega_k\delta(k)}{4\pi}
\right|_{-\infty}^{\infty}+\frac{1}{2}\int_{-\infty}^{\infty}\frac{dk}
{2\pi}\omega(k)\frac{d}{dk}\delta(k)\;.
\label{9}
\end{equation}
Note the appearance of a surface integration term. This term is not considered
in other treatments, see for example \cite{graham1}, \cite{goldhaber} and 
\cite{moden}.  
Eq. (\ref{9}) can be generalized to the case of potentials $U''[\phi_c(x)]$ 
that are not necessarily reflectionless. The expression is the same, we have 
only to give a generalized expression for the phase shift in terms of the 
$S$-matrix associated to the one dimensional scattering problem given by the 
continuous solutions of Eq. (\ref{3}). In terms of the reflection and 
transmission coefficient amplitudes, $R$ and $T$, the $S$-matrix reads
\cite{barton},
\begin{equation}
S(k)=\left(
\begin{array}{ll}
T(k)&-R^\ast(k)T(k)/T^\ast(k)\\
R(k)&T(k)\end{array}
\right)
\label{10}
\end{equation}
and the phase shift is given by \cite{barton}
\begin{eqnarray}
\delta(k)&=&\frac{1}{2i}\ln\det S(k)\nonumber\\
&=&\frac{1}{2i}\ln\left[\frac{T(k)}{T^\ast(k)}\right]\;,
\label{11}
\end{eqnarray}
where in passing to the second equality we have used the known property 
$|R|^2+|T|^2=1$. Since the $S$-matrix is unitary it is sure that $\delta(k)$
as given by Eq. (\ref{11}) is a real function of $k$. In the case of 
reflectionless potentials, since we have $T={\rm e}^{i\delta(k)}$, 
Eq. (\ref{11}) is trivially satisfied. 

The bare quantum mass corrections as given by Eq. (\ref{9}), is 
logarithmically divergent, since the phase shift behaves as $1/k$
for large $k$. Then, we have to renormalize such expression. In order to do
this we write Eq. (\ref{2}) in other equivalent form \cite{dunne} \cite{boya}
\begin{equation}
\Delta M_{bare}=\frac{1}{2}\int\frac{d\omega}{2\pi}{\rm Tr}\ln
\left[1+\frac{U''[\phi_c(x)]-m^2}{\omega^2-\frac{d^2}{dx^2}+m^2}\right]\;.
\label{12}
\end{equation}
The above expression is obtained with functional methods as the Euclidean
effective action per unit time evaluated at the static soliton configuration.
Eq. (\ref{12}) can be expanded in terms of Feynman graphs,
\begin{equation}
\Delta M_{bare}=\hspace{0.2cm}\parbox[b]{7.5cm}{ \raisebox{-1.2cm}
{\psfig{file=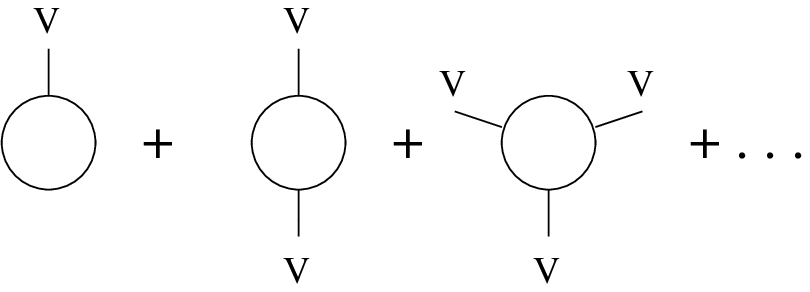,height=2.5cm,width=7.5cm}}}\quad ,
\label{12a}
\end{equation}
where the background field $V(x)= U''[\phi_c(x)]-m^2$. From the expansion
in Feynman graphs we note that the only divergent term is the tadpole graph,
\begin{equation}
\parbox[b]{2cm}{ \raisebox{-0.3cm}{\psfig{file=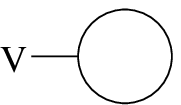,
height=0.85cm,width=1.55cm}}}\quad\hspace{-0.7cm}
= \frac{\left<V\right >}{4}\int_{-\infty}^\infty
\frac{dk}{2\pi}\frac{1}{\sqrt{k^2+m^2}}\;,
\label{13}
\end{equation} 
where 
\begin{equation}
\left<V\right>=\int_{-\infty}^\infty dx V(x)\;.
\label{14}
\end{equation}
As expected the tadpole graph is logarithmically divergent. Since Eqs. 
(\ref{9}) and (\ref{12a}) are equivalent, in order to regularize
the quantum mass correction as given by Eq. (\ref{9}), we 
subtract and add to it the tadpole graph given by Eq. (\ref{13}) obtaining
\begin{equation}
\Delta M_{bare}=
\left\{\frac{1}{2}\sum_i\omega_i-\left.\frac{\omega_k\delta(k)}{4\pi}
\right|_{-\infty}^{\infty}+\frac{1}{2}\int_{-\infty}^{\infty}\frac{dk}
{2\pi}\omega(k)\frac{d}{dk}\delta(k)-
\parbox[b]{2cm}{ \raisebox{-0.3cm}{\psfig{file=fig2.ps,
height=0.85cm,width=1.55cm}}}\quad\hspace{-0.7cm}\right\}+
\parbox[b]{2cm}{ \raisebox{-0.3cm}{\psfig{file=fig2.ps,
height=0.85cm,width=1.55cm}}}\quad\hspace{-0.7cm}\;.
\label{13a}
\end{equation}
In the above equation, the first part is finite and the full one loop
divergence is 
contained in the added tadpole graph only. Now we use the simplest
renormalization prescription that the added tadpole graph cancels exactly 
with adequate counterterms (that are present in the classical bare mass),
obtaining in this way unambiguously the renormalized soliton quantum mass 
correction at
one loop order. This renormalization prescription is 
equivalent to using a normal ordering prescription for the field operators,
that in 1+1 dimensions render finite any scalar field theoretical model 
\cite{cahill}. Then, we can identify the finite 
renormalized one loop soliton quantum mass correction, that we denote by
$\Delta M$, with the first part of Eq. (\ref{13a}),
\begin{equation}
\Delta M=\frac{1}{2}\sum_i\omega_i-\left.\frac{\omega_k\delta(k)}{4\pi}
\right|_{-\infty}^{\infty}+\frac{1}{2}\int_{-\infty}^{\infty}\frac{dk}
{2\pi}\omega(k)\frac{d}{dk}\delta(k)-
\frac{\left<V\right >}{4}\int_{-\infty}^{\infty}
\frac{dk}{2\pi}\frac{1}{\sqrt{k^2+m^2}}\;,
\label{15}
\end{equation}
where we used Eq. (\ref{13}) and it is understood that now
the parameters that appear in Eq. (\ref{15}) are the physical ones. In order
to compute the surface term in Eq. (\ref{15}) we only need to know the 
behavior of the phase shift for large $k$ and this is given by the  Born
approximation. The only term that contributes to the surface term is the first 
Born approximation, given by \cite{drazin}
\begin{equation}
\delta^1(k)=-\frac{\left<V\right>}{2k}\;.
\label{16}
\end{equation}
Replacing Eq. (\ref{16}) in Eq. (\ref{15}) we obtain,
\begin{equation}
\Delta M=\frac{1}{2}\sum_i\omega_i+\frac{\left<V\right>}{4\pi}+\frac{1}{2}
\int_{-\infty}^{\infty}\frac{dk}{2\pi}\omega(k)\frac{d}{dk}\delta(k)-
\frac{\left<V\right >}{4}\int_{-\infty}^\infty
\frac{dk}{2\pi}\frac{1}{\sqrt{k^2+m^2}}\;.
\label{17}
\end{equation}

In order to check that our formula given by the above equation is correct,
we use it to compute the one loop soliton quantum mass correction in the
sine-Gordon and $\phi^4$ kink models. \\

{\it The sine-Gordon Model}\par For this model the density potential is given
by
\begin{equation}
U(\phi)=\frac{m^4}{\alpha^2}\left[1-\cos\left(\frac{\alpha}{m}\phi\right)
\right]
\label{18}
\end{equation}
and $V(x)$ is given by
\begin{equation}
V(x)=-\frac{2m^2}{\cosh^2(mx)}\;.
\label{19}
\end{equation}
The potential $U''[\phi_c(x)]=V(x)+m^2$ admits only one discrete 
eigenvalue, $\omega_0^2=0$. The phase shift $\delta(k)=2\tan^{-1}(m/k)$ 
and from Eq. (\ref{14}) we obtain $\left<V\right>=-4m$. Subtituting these 
values in Eq. (\ref{17}) we obtain,
\begin{eqnarray}
\Delta M&=&-\frac{m}{\pi}-\frac{1}{2}\int_{-\infty}^{\infty} \frac{dk}{2\pi}
\frac{2m}{\sqrt{k^2+m^2}}+m\int_{-\infty}^{\infty}\frac{dk}{2\pi}\frac{1}
{\sqrt{k^2+m^2}}\nonumber\\
&=&-\frac{m}{\pi}\;.
\label{20}
\end{eqnarray}

{\it The $\phi^4$ kink model}\\
In this case the density potential is given by
\begin{equation}
U(\phi)=\frac{m^4}{32\alpha^2}\left[1-\frac{4\alpha^2}{m^2}\phi^2\right]^2\;,
\label{21}
\end{equation}
and $V(x)$ is given by
\begin{equation}
V(x)=-\frac{3m^2}{2\cosh^2(mx/2)}\;.
\label{22}
\end{equation}
In this case the potential $U[\phi_c(x)]$ admits two discrete eigenvalues, 
the zero mode eigenvalue $\omega_0=0$ and $\omega_1=\frac{m\sqrt{3}}{2}$.
The phase shift in this case is given by $\delta(k)=-2\tan^{-1}[3mk/
(m^2-2k^2)]$. From Eq. (\ref{14}) we obtain $\left<V\right>=-6m$. Replacing 
these values in Eq. (\ref{17}) we obtain,
\begin{eqnarray}
\Delta M&=&\frac{m\sqrt{3}}{4}-\frac{3m}{2\pi}-\frac{1}{2}
\int_{-\infty}^{\infty}\frac{dk}{2\pi}\frac{6m(2k^2+m^2)}
{(4k^2+m^2)\sqrt{k^2+m^2}}+\frac{3m}{2}\int_{-\infty}^{\infty}\frac{dk}{2\pi}
\frac{1}{\sqrt{k^2+m^2}}\nonumber\\
&=&\frac{m\sqrt{3}}{4}-\frac{3m}{2\pi}-\frac{3}{2}\int_{-\infty}^{\infty}
\frac{dk}{2\pi}\frac{m^3}{(4k^2+m^2)\sqrt{k^2+m^2}}\nonumber\\
&=&-m\left(\frac{3}{2\pi}-\frac{1}{4\sqrt{3}}\right)\;.
\label{23}
\end{eqnarray}
Note that Eqs. (\ref{20}) and (\ref{23}) are the same obtained previously,
with other methods. These results were obtained in Ref. \cite{neveu}. Then we 
can conclude that our formula, Eq. (\ref{17}), is correct. As we have mentioned 
previously, in the sine-Gordon and $\phi^4$ kink models, the potential $V(x)$ 
(or $U''[\phi_c(x)]$) is reflectionless. This property makes the calculation
easy. Next we use Eq. (\ref{17}) to compute the soliton quantum mass 
corrections in a model where $V(x)$ is not reflectionless.

{\it The model $\phi^2\cos^2\ln(\phi^2)$~~}
In Ref. \cite{flores} it has been introduced the model with density potential
given by
\begin{equation}
U(\phi)=\frac{1}{2}m^2B^2\phi^2\cos^2 \left[ \frac{1}{2B}\ln
\left(\frac{\alpha^2\phi^2}{9m^4}\right)\right]\;.
\label{24}
\end{equation}
This model has infinitely degenerate trivial vacua at the points 
\begin{equation}
\phi_n=\pm\frac{3m^2}{\alpha}\exp\left(\frac{2n+1}{2}\pi B\right)\;, 
~~n=0,\pm 1,\pm 2...
\label{25}
\end{equation}
The static soliton and anti-soliton solutions are given by \cite{flores}
\begin{equation}
\phi_c(x)=\pm\frac{3m^2}{\alpha}\exp\left[n\pi B
 \pm B\tan^{-1}\sinh(mx)\right]\;,~~n=0,\pm 1,\pm 2,..
\label{26}
\end{equation}
Using the above equation in the formula for $U''(\phi)$ obtained from Eq. 
(\ref{25}) we get
\begin{equation}
U''[\phi_c(x)]=m^2\left[1+\frac{(B^2-2)}{\cosh^2(mx)}
\pm 3B\frac{\tanh(mx)}{\cosh(mx)}\right]\;,
\label{27}
\end{equation}
from which we obtain for $V(x)$,
\begin{equation}
V(x)=m^2\frac{(B^2-2)\pm 3B\sinh(mx)}{\cosh^2(mx)}\;.
\label{28}
\end{equation}
The potential given by Eq. (\ref{27}) is exactly solvable \cite{cooper}. 
It has only one discrete eigenvalue, the zero mode solution $\omega_0=0$. 
Also the reflection and transmission coefficient amplitudes can be obtained 
exactly. For the transmission coefficient amplitude the result obtained is
\cite{khare}
\begin{equation}
T(k)=\frac{\Gamma(-1-ik/m)\Gamma(2-ik/m)\Gamma(1/2\mp iB-ik/m)
\Gamma(1/2\pm iB-ik/m)}{\Gamma(-ik/m)\Gamma(1-ik/m)\Gamma^2(1/2-ik/m)}\;,
\label{29}
\end{equation}
from which the phase shift $\delta(k)$ can be obtained using Eq. (\ref{11}).
Replacing Eq. (\ref{28}) in Eq. (\ref{14}) we get $\left<V\right>=2m
(B^2-2)$. Then using the preceding formulas in Eq. (\ref{17}), integrating by
parts and after some algebraic manipulations we obtain,
\begin{equation}
\frac{\Delta M}{m}=
-\frac{1}{\pi}-\frac{1}{2}\int_{-\infty}^{\infty} \frac{dq}{2\pi}\frac{q}
{\sqrt{q^2+1}}\left\{\frac{1}{2i}\ln\left[\frac{\Gamma(1/2-iB-iq)
\Gamma(1/2+iB-iq)\Gamma^2(1/2+iq)}{\Gamma(1/2+iB+iq)\Gamma(1/2-iB+iq)
\Gamma^2(1/2-iq)}\right]+\frac{B^2}{q}\right\}\;,
\label{30}
\end{equation}
where we have defined $q=k/m$. The integral in Eq. (\ref{30}) can not be
performed exactly, but for $B=0$ it vanishes and in this case we obtain,
$\Delta M=-m/\pi$, equal to the soliton quantum mass correction in the 
sine-Gordon Model. This is understood since for $B=0$ the potential $V(x)$
given by Eq. (\ref{28}) is equal to those given by Eq. (\ref{19}). For
other values of $B$ we can perform numerically the integration in Eq.
(\ref{30}). The result obtained is showed in Fig. \ref{fig1}. From Fig. 
\ref{fig1} we see that the soliton quantum mass correction is negative and 
decrease for increasing $B$. Although we do not compute exactly $\Delta M$ 
in this case, we call attention to the fact that all the steps have been done 
analytically and that only the final integration is done numerically. Then our 
computation, although not exact is precise. Frequently the sine-Gordon and 
$\phi^4$ kink models are used to test approximate or numerical methods 
developed to compute quantum corrections around static classical
configurations, see for example \cite{tom}. Here we present another model that 
can be used to this end. We believe that this model is more adequate to test 
approximate or numerical methods since the sine-Gordon and $\phi^4$ kink models
are very special: their quantum fluctuations are described by reflectionless 
potentials. 

I would like to remark that Eq. (\ref{17}) is equivalent to the one 
presented in Ref. \cite{graham1}, in which the authors have worked the 
case in which $V(x)$ is symmetric in $x$. 
The interested reader can prove this equivalence using the one-dimensional Levinson 
theorem. Although the final formula is equivalent the derivation is
different. The starting point in Ref. \cite{graham1} is 
Eq. (\ref{9}) without the surface term. Then the authors subtract the first Born 
approximation to the phase shift as given by Eq. (\ref{16}) in order to render 
finite the quantum mass correction. But, since the phase shift is 
divergent for $k=0$, they use (before doing the subtraction) the one
dimensional Levinson theorem \cite{barton}. By using the Levinson theorem
they cure the infrared divergence of the first Born approximation to the phase 
shift and also the surface term (not considered at the start) emerges. Finally
they use the same renormalization prescription that the added tadpole graph
cancels, obtaining in this way a formula that is equivalent to Eq.
(\ref{17}). Also in Ref. \cite{moden}, the surface term
in Eq. (\ref{9}) is not considered at the start but, it emerges after using
the called mode number cutoff. Finally I would like to call atention about 
the fact that if one consider in Ref. \cite{moden} as starting point Eq. (\ref{9}) 
with the surface term,  the energy-momentum cutoff 
regularization will give the correct answer for the soliton quantum mass corrections. 
I believe that the origin of part of the controversy raised in the literature about 
how to compute
correctly the soliton quantum mass corrections has been to use (incorrectly)
as starting point Eq. (\ref{9}) without the surface term. 
  \\\\\\
{\large \bf Acknowledgments}

I would like to thank N. Graham (UCLA) for very helpful discussions and to
A.P.C. Malbouisson (CBPF) for reading the manuscript. Also a grant from 
Conselho Nacional de Desenvolvimento Cient\'\i fico e Tecnol\'ogico 
(CNPq-Brazil) is acknowledged.


\begin{thebibliography}{99}
\bibitem{chan} L-H. Chan, Phys.Rev. Lett. {\bf 54}, 1222 (1985);
Phys. Rev. {\bf D55}, 6233 (1997).
\bibitem{graham1} N. Graham and R.L. Jaffe, Phys.Lett. {\bf B435}, 145 (1998);
arXiv:hep-th/9805150.
\bibitem{dunne} G. Dunne, Phys.Lett. {\bf B467}, 238 (1999); 
arXiv:hep-th/9907208.
\bibitem{goldhaber} A. S. Goldhaber, A. Litvintsev, P. Van Nieuwenhuizen,
 Phys.Rev. {\bf D64}, 045013 (2001); arXiv:hep-th/0011258.
\bibitem{bordag} M. Bordag, A.S. Goldhaber and P. Van Nieuwenhuizen and
D. Vassilevich, arXiv:hep-th/0203066. 
\bibitem{aragao} C. A. A. de Carvalho, Phys. Rev. {\bf D60}, 065021 (2002).
\bibitem{wimmer} A. Rebhan, P. van Nieuwenhuizen and R. Wimmer, 
arXiv:hep-th/0203137.
\bibitem{moden} A. Rebhan and P. van Nieuwenhuizen, Nucl.Phys. {\bf B508},
 449 (1997); arXiv:hep-th/9707163.
\bibitem{rajaraman} R. Rajaraman, {\it Solitons and Instantons},
North-holland Physics Publishing, Amsterdam (1982).
\bibitem{jackiw1} R. Jackiw, Rev. Mod. Phys. {\bf 49},  681 (1977).
\bibitem{barton} G. Barton, J. Phys. {\bf A}: Math. Gen. {\bf 18}, 479 (1985).
\bibitem{boya} L.J. Boya and J. Casahorran, Ann.Phys. {\bf 196}, 361 (1989).
\bibitem{cahill} K. Cahill, A. Comtet and R. Glauber, Phys.Lett. {\bf B64},
283 (1976).
\bibitem{drazin} P.G. Drazin and R.S. Johnson, {\it Solitons: An 
Introduction}, Cambridge University Press, England (1993).
\bibitem{flores} G. Flores-Hidalgo and N.F. Svaiter, 
Phys.Rev. {\bf D66}, 025031 (2002); arXiv:hep-th/0107043.
\bibitem{cooper} F. Cooper, A. Khare and U. Sukhatme, Phys.Rep. {\bf 251}, 267
(1995).
\bibitem{khare} A. Khare and U. Sukhatme, J.Phys.A:Math. Gen.{\bf 21}, L501
(1988).
\bibitem{neveu} R. Dashen, B. Hasslacher and A. Neveu, Phys.Rev. {\bf D10},
4114 (1974).
\bibitem{tom} C. Barnes and N. Turok, arXiv:hep-th/9711071; T. Weidig,
 arXiv:hep-th/9912005. 


 
\begin{figure}
\centerline {\epsfxsize=3.5in\epsfysize=2in\epsffile{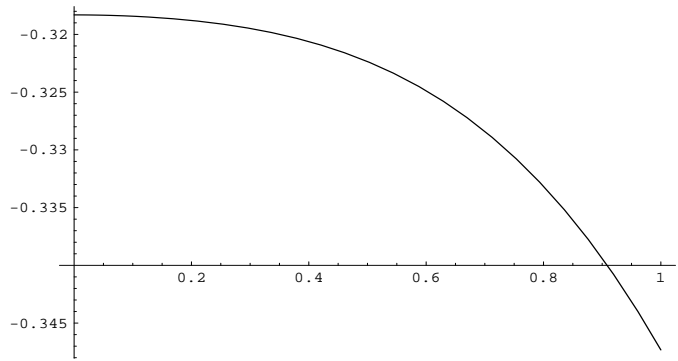} }
\caption{$\frac{\Delta M}{m}$ as function of $B$.}
\begin{picture}(10,10)
\end{picture}
\label{fig1}
\end{figure} 

\end{thebibliography}
\end{document}